\documentclass[aip, amsmath,amssymb]{revtex4-1}

\usepackage{graphicx}
\usepackage{dcolumn}
\usepackage{bm}
\usepackage{hyperref}

\usepackage[utf8]{inputenc}
\usepackage[T1]{fontenc}
\usepackage{mathptmx}
\usepackage{etoolbox}

\makeatletter
\def\@email#1#2{%
 \endgroup
 \patchcmd{\titleblock@produce}
  {\frontmatter@RRAPformat}
  {\frontmatter@RRAPformat{\produce@RRAP{*#1\href{mailto:#2}{#2}}}\frontmatter@RRAPformat}
  {}{}
}%
\makeatother
\begin{document}


\title{A two-color dual-comb system for time-resolved measurements of ultrafast magnetization dynamics using triggerless asynchronous optical sampling} 



\author{D. Nishikawa}
\author{K. Maezawa}
\author{S. Fujii}
\affiliation{Depertment of Physics, Faculty of Science and Technology, Keio University, 3-14-1 Hiyoshi, Kohoku-ku, Yokohama, Kanagawa 223-8522, Japan}
\author{M. Okano}
\affiliation{National Defense Academy, 1-10-20 Hashirimizu, Yokosuka, Kanagawa 239-8686, Japan}
\author{S. Watanabe}
\email[]{watanabe@phys.keio.ac.jp}
\affiliation{Depertment of Physics, Faculty of Science and Technology, Keio University, 3-14-1 Hiyoshi, Kohoku-ku, Yokohama, Kanagawa 223-8522, Japan}


\date{\today}

\begin{abstract}
We report on an Er-doped fiber (EDF)-laser-based dual-comb system that allows us to perform triggerless asynchronous optical sampling (ASOPS) pump-probe measurements of ultrafast demagnetization and spin precession in magnetic materials. Because the oscillation frequencies of the two frequency-comb light sources are highly stabilized, the pulse-to-pulse timing jitter is sufficiently suppressed and data accumulation without any trigger signals is possible. To effectively induce spin precession in ferromagnetic thin films, the spectral bandwidth of the output of one of the EDF frequency comb sources is broadened by a highly nonlinear fiber and then amplified at a wavelength of about 1030 nm by a Yb-doped fiber amplifier. The output of the other frequency comb source is converted to about 775 nm by second harmonic generation. We used this system to observe the ultrafast demagnetization and spin precession dynamics on the picosecond and nanosecond time scales in a permalloy thin film. This time-domain spectroscopy system is promising for the rapid characterization of spin-wave generation and propagation dynamics in magnetic materials.
\end{abstract}


\maketitle 

\section{Introduction}

Techniques that allow us to induce and control spin precession in magnetic materials are essential for spintronics~\cite{RevModPhys.76.323,maekawa2006concepts,Chumak2015}. For example, local spin precession can be used to induce magnetization reversal in magnetic storage devices, and therefore the development of methods to quickly induce local spin precession is important for future high-speed magnetic storage devices~\cite{PhysRevLett.79.1134,Stamps1999,Bauer2000,Acremann2001,Choi2001,Gerrits2001,Gerrits2002}. In addition, spin waves, which are propagating disturbances in the form of spin precession, have been studied intensively in terms of next-generation information carriers with low power consumption~\cite{Kostylev2005,stancil2009spin,Chumak2014}. To realize these applications related to spin precession and spin waves, it is important to elucidate the fundamental properties of magnetic materials and establish a method that enables us to quantitatively evaluate spin dynamics with high accuracy. Among the various possible evaluation methods, optical measurements have several advantages, because local spin precession can be easily induced by beam focusing and time-resolved measurements can be performed using pulsed lasers. Accordingly, the field of opto-spintronics has developed remarkably as ultrashort optical pulse laser technology has advanced~\cite{Kirilyuk2010,Hirohata2020}. Over the past 30 years, many experiments concerning laser-induced magnetization dynamics (including time-resolved measurements of ultrafast demagnetization~\cite{PhysRevLett.76.4250,PhysRevLett.78.4861,PhysRevLett.85.844} and the associated spin precession~\cite{PhysRevLett.82.3705,PhysRevLett.88.227201,Kimel2009,PhysRevB.85.184429}) have been performed using optical-pump-optical-probe techniques. In addition, it has been shown that the double-pulse excitation scheme can be used to control the spin precession amplitude~\cite{PhysRevB.73.014421,PhysRevB.74.012404,Kimel2007,Kampfrath2011,PhysRevB.97.014438} and to selectively excite higher-order spin precession~\cite{Okano2020}. Furthermore, spatial imaging of propagating spin waves has been achieved by scanning the beam spot across the sample~\cite{Satoh2012,Au2013,Iihama2016}. To accelerate the related research, a rapid and precise method for probing the temporal evolution of optically induced spin precession is indispensable. The typical spin precession period is in the range of 1 ps to 1 ns~\cite{Kirilyuk2010}, and the spin precession decay time after ultrafast demagnetization is on the order of several nanoseconds~\cite{PhysRevLett.76.4250,PhysRevLett.88.227201}. Therefore, to fully understand the spin precession dynamics, it is crucial to construct a rapid measurement system that possesses both a high time resolution and a long probe time in the time range from picoseconds to nanoseconds.

In conventional optical pump–probe setups, a relatively long time is needed to measure a single time trace because the delay time between the probe pulse and the pump pulse is controlled by an optical delay stage placed in the optical path. To overcome this disadvantage, the asynchronous optical sampling (ASOPS) method has been proposed, which requires no mechanical delay stage~\cite{Elzinga1987,Adachi1995,Yasui2005,Bartels2007,Stoica2008,Abbas2014,Good2015,Krauss2015,Hitachi2018,Scherbakov2019,Asahara2020,Nakagawa2022,Okano2022}. Because of the different repetition frequencies, the time interval between the probe pulse and pump pulse is a (periodic) function of time. Subsequent probe pulses detect the sample condition (e.g., the instantaneous direction of the magnetization) at different times after the pump pulse irradiation. When the time interval between the probe pulse and the pump pulse becomes equal to the pump pulse period, the effective delay time is reset to zero and the above scanning procedure is repeated. In this way, the delay of the probe pulse occurs automatically and the temporal resolution is determined by the repetition rates of the two pulse trains. As a result, the acquisition time of the time-resolved measurement can be significantly reduced, and this enables us to greatly improve the signal-to-noise ratio by accumulating the signal~\cite{Bartels2007}. 

The ASOPS method has already been used in time-resolved measurements of various materials, and it has also been applied to terahertz time-domain spectroscopy~\cite{Yasui2005,Good2015,Nakagawa2022,Okano2022} and ultrasonic measurements~\cite{Stoica2008,Hitachi2018}. The investigation of optically induced magnetization precession using ASOPS has also been reported~\cite{Scherbakov2019}. However, despite the above-mentioned automatic delay in an ASOPS system, an electronic trigger signal is usually required to determine the time origin of each temporal profile in order to accumulate the data and improve the signal-to-noise ratio (this trigger is required in the case of an incomplete stabilization of the repetition frequencies of the two pulse trains, which causes a fluctuation in the emission timing). The limited time resolution of the trigger signal may lead to a timing jitter, which can also cause a reduction in the signal intensity of the averaged time-resolved waveform~\cite{Asahara2020}. To reduce the timing jitter, the use of a fully frequency-stabilized dual-comb light source is considered to be a good solution. Recently, Asahara {\it et al.} reported a dual-comb system for the ASOPS-based characterization of the carrier lifetime in InAs quantum dots, and showed that a time-resolved pump–probe measurement with a high stability and an extremely wide temporal dynamic range can be realized~\cite{Asahara2020}. We used a dual-comb system to perform ASOPS-based terahertz time-domain spectroscopy measurements, and demonstrated that data accumulation over 48 minutes in a triggerless manner is possible without signal degradation~\cite{Nakagawa2022}. These studies were possible due to developments in the area of fully frequency-stabilized frequency combs using Er-doped fiber (EDF) laser systems~\cite{Inaba2006}. However, additional improvements are required to apply the EDF-laser-based characterization of ultrafast spin precession to a wider range of materials. Firstly, a wavelength conversion to shorter wavelengths is required to tightly focus the pump and probe pulses on the sample surface to increase the excitation density and improve the spatial resolution in imaging applications. Secondly, the wavelengths of the pump and probe beams should be different to be able to separate them by a wavelength filter. Finally, a significant increase in the pump power may be required to strongly excite the sample (the typical output power of an EDF laser system is $\sim$100 mW). To obtain shorter wavelengths, second harmonic generation (SHG) of one of the frequency combs should be a good solution, but another conversion mechanism needs to be used for the other frequency comb to satisfy the above three requirements.  In 2017, Liu {\it et al.} proposed a Yb-doped fiber amplifier with a spectrally broadened EDF-laser-based frequency comb system as the seed light source~\cite{Liu2017,Liu2017a}.They achieved an output power of up to 356~mW at a wavelength of $\sim$1030 nm. Using such a light source should enable us to fulfill the above three requirements. 

In this paper, we report on a two-color EDF-laser-based dual-comb system for optical pump–probe measurements of ultrafast demagnetization and spin precession in magnetic materials. The wavelength of the EDF frequency comb source for the pump pulses was converted to about 1030 nm using an Yb-doped fiber amplifier, and pulse compression by a grating pair resulted in pump pulses with a temporal width of $\sim$0.2 ps and a pump power of 15 mW, which is sufficiently high to induce spin precession. The wavelength of the frequency comb source for the probe pulses was converted to about 775 nm by SHG. By using this two-color frequency comb system for optical pump–probe measurements, we observed ultrafast demagnetization and the resulting spin precession in a permalloy thin film sample. We show that the experimentally determined dependence of the precession frequency on the external magnetic field agrees well with the theoretical prediction. The measurement speed of this ASOPS system is much faster than the measurement speeds of conventional setups using mechanical delay stages, and therefore our approach is expected to be highly useful for studies on ultrafast magnetization dynamics.

\section{SYSTEM COMPONENTS}
\subsection{Frequency comb sources with ultrahigh-precision frequency stabilization} \label{sec:stab}
\begin{figure}
\includegraphics[width=0.9\linewidth]{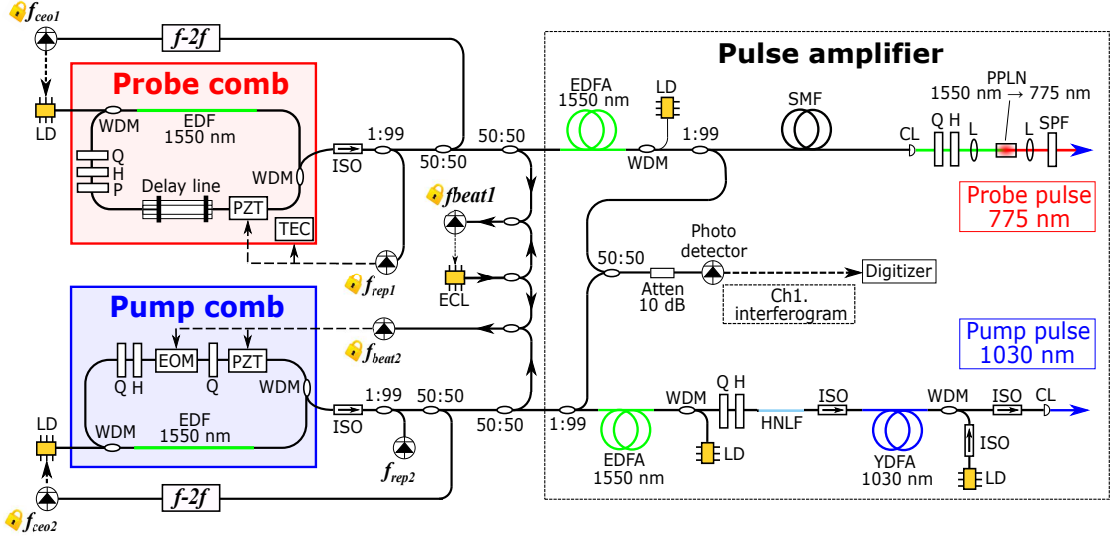}
\caption{\label{fig1_dual_comb} Schematic of the two-color dual-comb system. The components are as follows. ECL: external cavity laser, LD: laser diode, EDF: Er-doped fiber, EDFA: Er-doped fiber amplifier, YDFA: Yb-doped fiber amplifier, HNLF: highly nonlinear fiber, PZT: piezoelectric element, TEC: thermoelectric cooler, EOM: electro-optic phase modulator, WDM: wavelength division multiplexing coupler, Q: quarter-wave plate, H: half-wave plate, P: polarizer, ISO: isolator, Atten: attenuator, $f$-2$f$: $f$-2$f$ interferometer, SMF: single-mode fiber, PPLN: periodically-poled lithium niobate, SPF: short-pass filter, L: lens, CL: collimator lens, and Ch1.: Channel 1.}
\end{figure}

Figure~\ref{fig1_dual_comb} shows a schematic diagram of the dual-comb light source used for our ASOPS system. To realize a tight stabilization of the repetition frequencies of the pump comb and the probe comb, we tightly lock all individual frequency components of the two combs in the optical frequency region. We use the same frequency-stabilization method as that reported by Fukuda {\it et al.}~\cite{Fukuda2021}. Firstly, the carrier-envelope offset frequencies of the probe comb and pump comb, denoted by $f_\mathrm{ceo1}$ and $f_\mathrm{ceo2}$, respectively, are detected by an $f$–2$f$ self-referencing interferometer~\cite{Holzwarth2000,Jones2000} and locked to radio frequency (RF) reference signals from two function generators (WF1968 and WF1974, NF Corp.) via feedback control of the laser-diode current for each of the two pump laser diodes (LDs) (1999CB and 1999CHB, 3SP Technologies) by analog proportional-integral control circuits. Secondly, the repetition frequency of the probe comb, $f_\mathrm{rep1}$, is also locked to an RF reference signal. The fiber length is controlled by a piezoelectric element (PZT) and by changes in the ambient temperature through a thermoelectric cooler (TEC) (TEC1-12708), which also changes the refractive index~\cite{Nakajima2010,Asahara2017}. As a result, all optical frequency components of the probe comb are stable with respect to the RF signals and obey the relation $f_\mathrm{probe}(n) = f_\mathrm{ceo1} + n\cdot f_\mathrm{rep1}$, where $n$ is an integer.  Finally, for each frequency comb, we need to stabilize the optical beat frequency between one of the comb lines and a continuous-wave narrow-band external cavity laser (ECL) (Redfern Integrated Optics, PLANEX) with a wavelength of 1550.11 nm and a line-width of 1.6 kHz (the beat frequencies obtained for the probe and pump comb are denoted by $f_\mathrm{beat1}$ and $f_\mathrm{beat2}$, respectively)~\cite{Fukuda2021}. The frequency of the ECL, $f_\mathrm{cw}$, is stabilized by locking $f_\mathrm{beat1}$ to an RF reference signal via feedback control of the ECL current to fulfill the relation $f_\mathrm{cw} = f_\mathrm{probe}(n_\mathrm{cw1}) + f_\mathrm{beat1}$, where $f_\mathrm{probe}(n_\mathrm{cw1})$ is the frequency of the probe-comb component adjacent to $f_\mathrm{cw}$ and $n_\mathrm{cw1}$ is the corresponding integer. Then, one of the optical frequencies of the pump comb is stabilized with $f_\mathrm{beat2}$ by controlling the voltage applied to an intracavity electro-optic phase modulator (EOM) and the piezoelectric element.  The repetition frequency of the pump comb, $f_\mathrm{rep2}$, is indirectly determined by the relation $f_{cw} = f_\mathrm{ceo2} + n_\mathrm{cw2}\cdot f_\mathrm{rep2} + f_\mathrm{beat2}$, where $n_\mathrm{cw2}$ is another certain integer. We adopted a global positioning system (GPS)-controlled rubidium clock with a relative uncertainty on the order of $10^{-12}$ as the master clock that provides the reference signal for the function generators. Thus, all optical phases of the two frequency combs are synchronized, resulting in a fully stabilized dual-comb system. 

In this study, we set $f_\mathrm{beat2}$ to a value that results in
\begin{equation}
    f_\mathrm{rep2} = \frac{N+1}{N} f_\mathrm{rep1},
\end{equation}
where $N$ is an integer that defines the number of data points of a single scan of the ASOPS system. If we define $\Delta f_\mathrm{rep}$ as the difference between the repetition frequencies of the two combs, $f_\mathrm{rep2}-f_\mathrm{rep1}$, $N$ can be written as
\begin{equation}
    N = \frac{f_\mathrm{rep1}}{\Delta f_\mathrm{rep}}. \label{eq:N}
\end{equation}
Regarding the time-domain description of the measurement, we need to consider the delay time of each probe pulse with respect to the pump pulse that reaches the sample immediately before the probe pulse: the delay times of subsequent probe-comb pulses increase in steps of $\tau_\mathrm{step} = \Delta f_\mathrm{rep}/(f_\mathrm{rep1} \cdot f_\mathrm{rep2})=1/(N \cdot f_\mathrm{rep2})$~\cite{Asahara2020}. Furthermore, we record the signal probed by the probe-comb beam with a sampling frequency of $f_\mathrm{rep1}$.  Therefore, each data point of the recorded signal represents the sample condition at a time $\tau_\mathrm{step}$ later than that of the previous data point. Furthermore, the total delay time in this implementation of the ASOPS method is $\tau_\mathrm{total} = N\times \tau_\mathrm{step} = 1 / f_\mathrm{rep2}$. The temporal dynamic range~\cite{Asahara2020}, i.e., the ratio between the total delay time and the delay-time increment, 
\begin{equation}
    \tau_\mathrm{total}/\tau_\mathrm{step} = N.
\end{equation}
Note that the actual measurement time required to record a single scan in the laboratory frame is $\tau_\mathrm{scan} = 1 / \Delta f_\mathrm{rep}$.

\subsection{Wavelength conversion, pulse amplification, and pulse compression} \label{sec:pulse}
\begin{figure}
\includegraphics[width=0.9\linewidth]{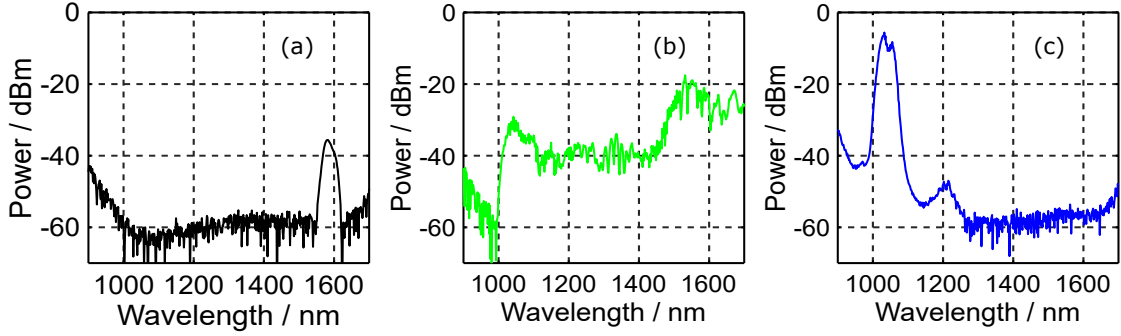}
\caption{\label{fig2_optical_power} Spectra of the pump-comb pulse (a) after the HNLF without amplification in the EDFA, (b) after the HNLF with amplification in the EDFA, and (c) after the YDFA with amplification in the EDFA. The output power of the amplified pump-comb beam after the HNLF was 186 mW, and that after the YDFA was 105 mW.}
\end{figure}

The right-hand side of Fig.~\ref{fig1_dual_comb} shows a schematic diagram of the wavelength conversion and pulse amplification in our dual-comb system (the pulse compression is explained later). The 1550-nm output of the pump-comb source is first amplified by an Er-doped fiber amplifier (EDFA) using an Er-doped fiber as the amplification medium (ER30-4/125, Thorlabs), and then the spectral bandwidth is broadened to nearly 1000 nm by a highly nonlinear fiber (HNLF). To clarify the spectrum of the pump pulse used in our experiment, Figs.~\ref{fig2_optical_power}(a) and \ref{fig2_optical_power}(b) show the optical spectra of the pump pulse after the HNLF without and with the amplification by the EDFA, respectively (we can confirm that the center wavelength of the seed pulse was 1580.0 nm and the spectral bandwidth was about 60 nm in the case of no amplification). The amplification by the EDFA leads to a significant broadening of the spectrum of the seed pulse. As shown in Fig.~\ref{fig1_dual_comb}, the broadened pulse passes through a fiber isolator (IO-F-1030, Thorlabs) and is used as a seed pulse for a second amplification stage: a Yb-doped fiber amplifier (YDFA) (amplification medium: SCF-YB550-4/125-19, CorActive) is used to amplify the spectral component around 1030 nm. In our experiment, the exact position of the spectral peak after the YDFA was 1032.8 nm as shown in Fig.~\ref{fig2_optical_power}(c). The output power of the amplified pump pulse was around 100 mW, which is sufficiently high for the experiments in this work. The 1550-nm output of the probe-comb source is also amplified by an EDFA (ER30-4/125, Thorlabs), and then the center wavelength is converted to about 775 nm by SHG in a periodically-polled lithium niobate (PPLN) crystal (in our experiment, the obtained center wavelength was 779.4 nm). Thus, we achieved a two-color dual-comb system with fully stabilized emission timings.

\begin{figure}
\includegraphics[width=0.6\linewidth]{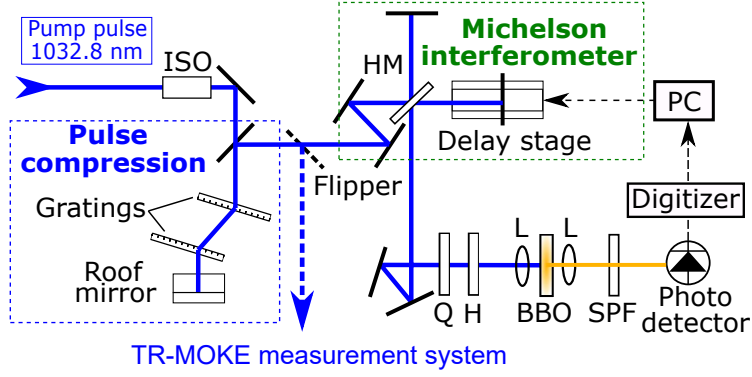}
\caption{\label{fig3_pulse_compresiion} Schematic of the pulse compression stage and the pulse-width measurement system. The components are as follows. ISO: isolator, HM: half mirror, Q: quarter-wave plate, H: half-wave plate, BBO: barium borate crystal, SPF: short-pass filter, L: lens, and TR-MOKE: time-resolved magneto-optical Kerr effect.}
\end{figure}

As a result of positive dispersion in the EDF and the Yb-doped fiber used for amplification, the amplified optical pulses are positively chirped. Such a temporal broadening of the pulse leads to a reduction in the time resolution of the optical pump–probe measurement and a reduction in the SHG efficiency. Therefore, it is necessary to compensate for the dispersion by either providing a negative chirp or pulse compression. Figure~\ref{fig3_pulse_compresiion} is used to describe the pulse compression and the optics used to characterize the pump pulse. The output of the YDFA is first compressed by two parallel-aligned transmission gratings (T-1000-1040-3212-95, LightSmyth) with a line density of 1000 lines/mm. The distance between the two gratings was about 31 mm and the angle of incidence was about 28$^\circ$. The optimization of the distance between the two gratings was performed by monitoring the pulse width after pulse compression using the interferometric autocorrelation measurement shown on the right-hand side of Fig.~\ref{fig3_pulse_compresiion}. This system consists of a Michelson interferometer, a beta-barium borate (BBO) crystal for SHG, and the required components for detection. After the optimization, a mirror was inserted into the beam path by a mechanical flipper to redirect the compressed pump-comb beam to the sample. Regarding the probe pulse, dispersion compensation is realized by a single-mode fiber (SMF) with negative dispersion (SMF-28, Thorlabs) in front of the SHG optics as shown in Fig.~\ref{fig1_dual_comb}. The pulse-width optimization of the seed pulse for SHG was performed by monitoring the pulse width of the seed pulse with another interferometric autocorrelation measurement system and changing the length of the SMF. The probe pulse used in the experiments (779.4 nm) was generated by the SHG of the optimized seed pulse.

\begin{figure}
\includegraphics[width=0.5\linewidth]{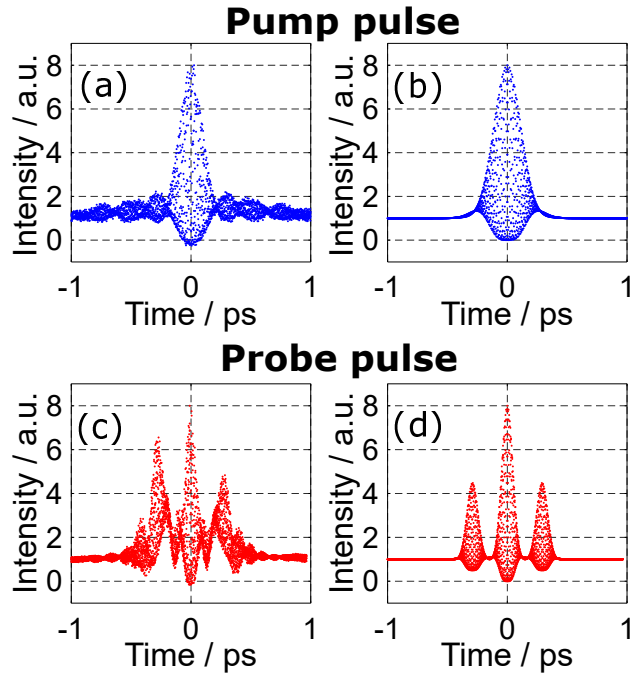}
\caption{\label{fig4_pulse_width} (a) The interferometric autocorrelation signal of the pump pulse and (b) the result obtained by fitting the experimental data to Eq. (\ref{eq:G2}). The experimental data is normalized (the background signal is 1 and the maximum signal value is 8). The fitted parameters of the pump light were $\tau_g=0.205$ ps, $a=1.72$, and $f = 277.0$ THz. (c) The interferometric autocorrelation signal of the probe pulse and (d) the result obtained by fitting the experimental data to Eq.~(\ref{eq:seed}). The probe light was a double pulse, and we estimated $\tau_g=0.0707$ ps, $a = 1.03$, $f = 194.3$ THz, $d = 0.291$ ps, and $A = 0.953$.}
\end{figure}

The time resolution of our system, which is discussed in Sec.~\ref{sec:fac}, also depends on the electric field envelope of the pulses. The interferometric autocorrelation signal of the compressed pump pulse is shown in Fig.~\ref{fig4_pulse_width}(a). To fit the data, we assume that the pump pulse can be described by a linearly chirped Gaussian pulse, i.e., the complex electric field envelope of the pump pulse, $\tilde{\varepsilon}_\mathrm{pump}(t)$, can be written as follows:

\begin{equation}
\tilde{\varepsilon}_\mathrm{pump}(t) \propto \exp{\left( -(1+ia)\left( \frac{t}{\tau_g}\right)^2 
 \right)},
\end{equation}
where $t$ is the time, $\tau_g$ is the pulse width, and $a$ is the chirp parameter. In this case, the interferometric autocorrelation signal $G_2$ should become~\cite{Diels2006}:

\begin{equation}
\begin{split}
G_2(\tau) \propto 1 +2 \exp{\left( -\left( \frac{\tau}{\tau_g}\right)^2 
 \right)} &+4 \exp{\left( -\frac{a^2+3}{4}  \left( \frac{\tau}{\tau_g}\right)^2 \right)} \cos{\left( \frac{a}{2}  \left( \frac{\tau}{\tau_g}\right)^2 \right)} \cos{(2 \pi f_0\tau)}\\
 &+4 \exp{\left( -(a^2+1)  \left( \frac{\tau}{\tau_g}\right)^2 \right)}\cos{(4 \pi f_0\tau)}, \label{eq:G2}
 \end{split}
\end{equation}
where $\tau$ is the delay time, and $f_0$ is the center frequency of the pump pulse. We fitted the experimental data in Fig.~\ref{fig4_pulse_width}(a) to Eq.~(\ref{eq:G2}) by the least squares method, and Fig.~\ref{fig4_pulse_width}(b) shows the fitting result ($\tau_g=0.205$ ps, $a = 1.72$, and $f_0 = 277.0 $ THz). 

To discuss the electric field envelope of the probe pulse, we analyze the autocorrelation signal of the ``seed'' pulse (with a wavelength of 1550 nm) instead of that of the probe pulse generated by SHG, because we were not able to directly evaluate the pulse width of the SHG output due to the small intensity. The interferometric autocorrelation signal of the dispersion-compensated seed pulse is shown in Fig.~\ref{fig4_pulse_width}(c). Several peaks are observed, and this signal cannot be reproduced by the interferometric autocorrelation signal for a single pulse as described in Eq.~(\ref{eq:G2}). This multiple-peak structure originates from a seed pulse with a double-peak structure, which may be due to the peak power clamping effect that sometimes occurs in mode-locked laser systems~\cite{Tang2005}. To account for the multiple-peak structure in the fitting process, we calculated the autocorrelation signal for the case that the complex electric field amplitude of the seed pulse, $\tilde{\varepsilon}_\mathrm{seed}(t)$, is a double Gaussian pulse;
\begin{equation}
\tilde{\varepsilon}_\mathrm{seed}(t) \propto \exp{\left( -(1+ia)\left( \frac{t}{\tau_g}\right)^2 
 \right)}+ A \exp{\left( -(1+ia)\left( \frac{t-d}{\tau_g}\right)^2 
 \right)}, \label{eq:seed}
\end{equation}
where $d$ is the time difference between the two pulses and $A$ is the ratio of the electric field amplitudes of the two pulses. Here, the autocorrelation function of the seed pulse can be derived in the same manner as the pump pulse for numerical fitting. Figure~\ref{fig4_pulse_width}(d) shows the obtained fitting result ($\tau_g = 0.0707$ ps, $a = 1.03$, $f_0 = 194.3$ THz, $d = 0.291$ ps, and $A = 0.953$), which reproduces the experimentally observed signal well. The difference between the heights of the left and right peaks in Fig.~\ref{fig4_pulse_width}(c) is due to an inaccurate spatial overlap of the two beams due to a misalignment of the Michelson interferometer. As mentioned above, the second harmonic of this seed pulse was used as the probe pulse in the experiment. Although we were not able to directly measure the autocorrelation signal of the SHG output, the numerical prediction of the pump–probe signal provided in Sec.~\ref{sec:fac} shows that the temporal broadening of the two peaks of the seed pulse due to the dispersion of the PPLN crystal was not larger than the temporal separation between the two peaks, and therefore the temporal separation $d$ has a strong influence on the pump–probe signal.

\subsection{ASOPS-based TR-MOKE measurement}

\begin{figure}
\includegraphics[width=0.5\linewidth]{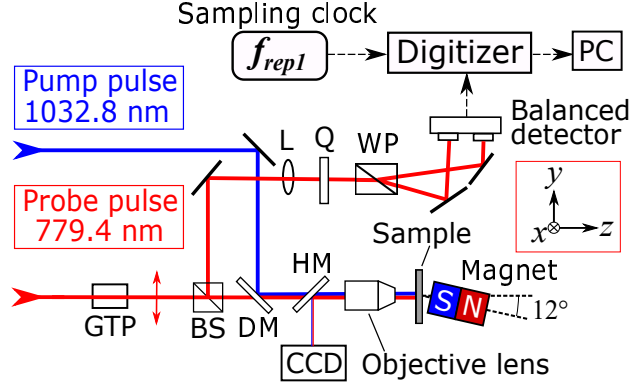}
\caption{\label{fig5_TR-MOKE} Schematic of the ASOPS-based TR-MOKE measurement. The components are as follows. GTP: Glan–Thompson prism, BS: beam splitter, DM: dichroic mirror, HM: half mirror, CCD: charge-coupled-device camera, Q: quarter-wave plate, L: lens, and WP: Wollaston prism.}
\end{figure}

Figure~\ref{fig5_TR-MOKE} shows a schematic of the time-resolved magneto-optical Kerr effect (TR-MOKE) measurement performed by using our ASOPS system. The blue curve represents the compressed pump-comb pulse, which is reflected by a dichroic mirror and focused on the sample by an objective lens with a magnification of 50$\times$, a working distance of 8.2 mm, and a numerical aperture of 0.55 (in our experiment, the spot size and the excitation fluence of the 1032.8-nm pump-comb beam on the sample surface were 2.18~\textmu m and 6.80~$\mathrm{mJ/cm^2}$, respectively, where the former value was estimated by the knife-edge method). The red curve represents the probe pulse, which is aligned coaxially with the pump pulse and also focused on the magnetic sample using the same objective lens (the spot size and the excitation fluence of the 779.4-nm probe-comb beam on the sample surface were 1.61~\textmu m and 10.7 $\mathrm{mJ/cm^2}$, respectively). The spatial overlap between the pump-comb beam and the probe-comb beam can be monitored by a charge-coupled-device camera (IC4133BU, Gazo). After passing through the Glan-Thompson prism (GTP) shown on the left-hand side of Fig.~\ref{fig5_TR-MOKE}, the probe-comb pulse is linearly polarized along the y-direction, and the polarization plane rotates when it is reflected from the sample surface due to the polar Kerr effect. The reflected probe pulse is then deflected by a beam splitter, transmitted through a quarter-wave plate, and split by a Wollaston prism. The balanced detection scheme is used to analyze the information contained in the split beam with a frequency resolution of $f_\mathrm{det} = 1$ MHz (PDB210A, Thorlabs). The signal is recorded by a digitizer (M2p.5962-x4, Spectrum) with a sampling frequency of $f_\mathrm{rep1}$. The time-domain waveform of the TR-MOKE signal is automatically obtained as explained in Sec.~\ref{sec:stab}; a single TR-MOKE scan contains $N$ sampling points, where $N$ is determined by Eq.~(\ref{eq:N}). This system does not need to use a trigger signal to repeatedly start the recording of individual waveforms for averaging, because it has a tight frequency stabilization of the two pulses~\cite{Okano2022}.

Regarding the used sample, we prepared a 20-nm-thick permalloy ($\mathrm{Ni_{0.8}Fe_{0.2}}$) film deposited on a silicon substrate by vacuum electron beam deposition. We applied an external magnetic field to the sample by a neodymium magnet, where the angle between the magnet axis and the sample normal direction was 12$^\circ$. The magnitude of the magnetic field, $H_\mathrm{ext}$, was controlled by changing the distance between the sample and the magnet and calibrated by a Gauss meter (TM-701, KANETEC) before the experiment. 

\subsection{Factors that determine the time resolution of the ASOPS system} \label{sec:fac}

\begin{figure}
\includegraphics[width=0.4\linewidth]{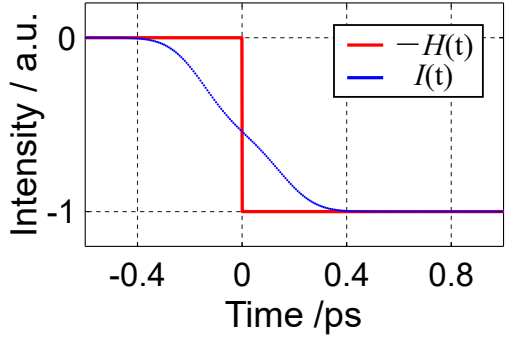}
\caption{\label{fig6_time_resolution} Temporal profile of the response of our system to the Heaviside step function calculated using Eq.~(\ref{eq:I}). The duration of the rise/fall of the signal is about 0.8 ps.}
\end{figure}

There are two factors that limit the time resolution; the pulse widths of the pump and probe pulses, and the detector bandwidth. The effect of the pulse width is considered first: The approximate shape of the step response of our system can be determined by calculating the following temporal profile of the signal $I(t)$~\cite{Polli2010};

\begin{equation}
I(t) \propto \int_{-\infty}^{\infty} d \tau'  \int_{-\infty}^{\infty} d \tau''  ~\tilde{\varepsilon}_\mathrm{pump}^2(\tau') \cdot -\mathcal{H}(t-\tau'+\tau'') \cdot \tilde{\varepsilon}_\mathrm{probe}^2(\tau''), \label{eq:I}
\end{equation}
where $\tilde{\varepsilon}_\mathrm{pump}(\tau')$ and $\tilde{\varepsilon}_\mathrm{probe}(\tau'') (\propto \tilde{\varepsilon}_\mathrm{seed}^2(\tau''))$ are the complex electric field amplitudes of the pump and probe pulses determined in Sec.~\ref{sec:pulse}, respectively. $\mathcal{H}(t)$ is the Heaviside step function. Here, we neglect the temporal broadening of each peak of the double-peak pulse in the PPLN crystal, and therefore the parameter $a$ in Eq.~(\ref{eq:seed}) for the seed pulse is also used to describe the pulse generated by SHG. Figure~\ref{fig6_time_resolution} shows the temporal profile of $I(t)$. The rise time of $I(t)$ is about 0.8 ps, which is the approximate time resolution of our system if the detector bandwidth is not the limiting factor. We found that the temporal separation $d$ in Eq.~(\ref{eq:seed}) has a strong influence on the rise time, and therefore it is important to suppress multiple-peak structures in the pump and probe pulses to improve the time resolution.

Regarding the impact of the detector bandwidth, because the measurement time required to obtain a single time-domain waveform is very short in an ASOPS system, the detector bandwidth $f_\mathrm{det}$ ($\sim$1 MHz in our system) sometimes limits the time resolution of the pump–probe signal. As explained in Sec.~\ref{sec:stab}, the actual required measurement time of a single waveform, $\tau_\mathrm{scan}$, is the inverse of $\Delta f_\mathrm{rep}$, and the maximum time delay of the waveform is the inverse of $f_\mathrm{rep2}$ ($\sim$61.5 MHz in our system). The time resolution determined by the detector bandwidth is thus $t_\mathrm{det}=\Delta f_\mathrm{rep}/(f_\mathrm{det} \cdot f_\mathrm{rep2})$~\cite{Okano2022}. Therefore, $t_\mathrm{det}$ becomes shorter as $\Delta f_\mathrm{rep}$ becomes smaller. In our system, $t_\mathrm{det}=$ 0.8 ps for $\Delta f_\mathrm{rep}=49$ Hz. Therefore, the time resolution is limited by the detector bandwidth if $\Delta f_\mathrm{rep}>49$ Hz, and it is determined by the pulse widths if $\Delta f_\mathrm{rep}<49$ Hz is used.

\section{RESULTS AND DISCUSSION}
\subsection{Observation of ultrafast spin precession}
\begin{figure}
\includegraphics[width=0.8\linewidth]{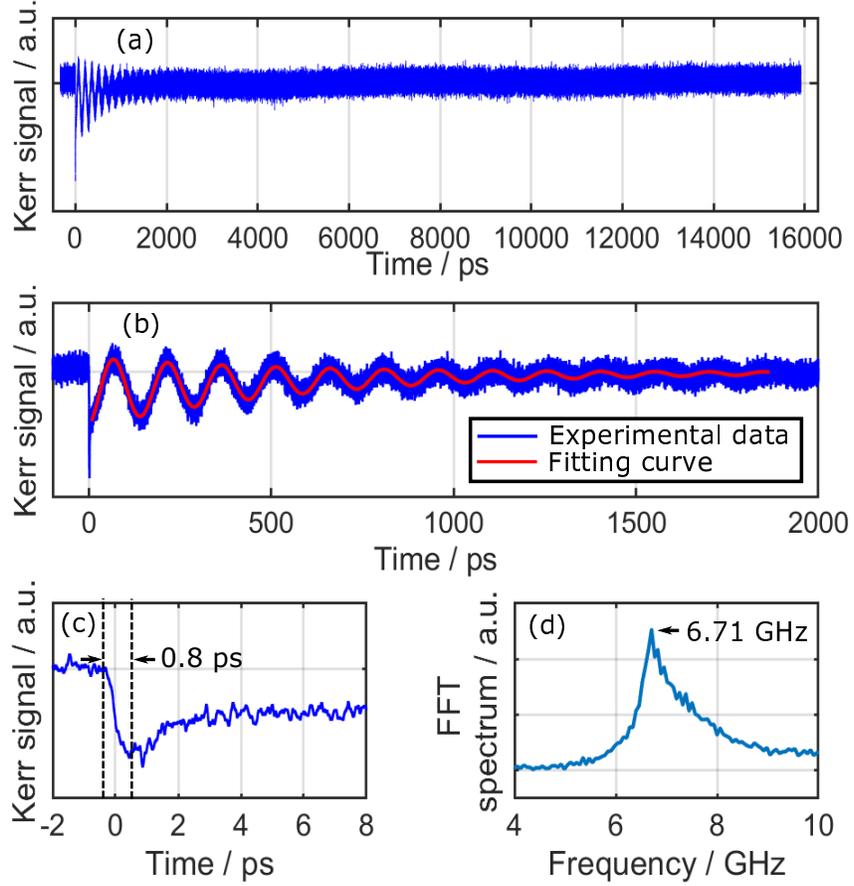}
\caption{\label{fig7_spin_precession} (a) The result of the ASOPS-based ultrafast time-resolved measurement of the magnetic Kerr rotation. (b) The TR-MOKE data in the range where spin precession occurs. The fitting result is shown by the red curve with the following parameters: $f_r = 6.73$ GHz, $\tau_a = 592$ ps, and $\tau_b = 673$ ps. (c) The TR-MOKE data in the range where ultrafast demagnetization occurs. (d) The FFT spectrum of the data shown in (a).}
\end{figure}

The developed ASOPS pump–probe system was first used to characterize the magnetization dynamics induced by optical excitation of the permalloy (Py) thin film in a relatively strong magnetic field. Table~\ref{tab:table1} shows the measurement conditions for this experiment. We set $\Delta f_\mathrm{rep}\approx25.140$~Hz, and therefore the time resolution is limited by the pulse widths of the pump and probe pulses. Figure~\ref{fig7_spin_precession}(a) shows the TR-MOKE signal obtained by averaging 1000 consecutive scans, and the time range of a single scan is $\tau_\mathrm{total} = 1/ f_\mathrm{rep2} = 16.25$ ns. The total measurement time required to record this data was 39.78 s. Figure~\ref{fig7_spin_precession}(b) shows a magnification of the data in Fig.~\ref{fig7_spin_precession}(a) around the time origin. We can see that the optically-induced spin precession lasts for about 2.0 ns. We evaluated the oscillation parameters of the spin precession by fitting the experimental data to the following function~\cite{PhysRevB.85.184429}:
\begin{equation}
I_\mathrm{Kerr}(t) \propto B \exp{\left(-\frac{t}{\tau_a}\right) 
}\cos{(2 \pi f_r t +\theta)} + C \exp{\left(-\frac{t}{\tau_b}\right) 
}. \label{eq:kerr}
\end{equation}
Here, the first term corresponds to the magnetization oscillation component and the second term corresponds to the nonmagnetic background component. $B$ and $C$ are the amplitudes of the two components, $f_r$ is the resonance frequency, $\theta$ is the phase, and $\tau_a$ and $\tau_b$ are damping constants. The approximate values of $f_r$, $\tau_a$, and $\tau_b$ for the blue curve in Fig.~\ref{fig7_spin_precession}(b) are 6.73 GHz, 592 ps, and 673 ps, respectively. These values are on the same order of magnitude as those in the literature~\cite{PhysRevLett.79.1134,Mizukami2001}. Figure~\ref{fig7_spin_precession}(c) shows a magnified view of the signal within the first eight picoseconds after the pump-pulse irradiation: the TR-MOKE signal decreases sharply within about 0.8 ps and then slightly recovers, indicating ultrafast demagnetization due to the pump-pulse irradiation. 
In a previous study on ultrafast demagnetization in ferromagnetic nickel, a demagnetization time of about 100 fs was observed~\cite{PhysRevLett.76.4250}. The time duration of the initial sharp decrease in Fig.~\ref{fig7_spin_precession}(c) is significantly longer. We consider that the observed duration of 0.8 ps is consistent with the time resolution of our system as determined in Sec.~\ref{sec:fac}. Figure~\ref{fig7_spin_precession}(d) shows the spectrum obtained by the fast Fourier transform (FFT) of the TR-MOKE signal in Fig.~\ref{fig7_spin_precession}(a). The peak is observed at 6.71 GHz, which is close to the value obtained by fitting the signal in Fig.~\ref{fig7_spin_precession}(b) to Eq.~(\ref{eq:kerr}).

\begin{table}
\caption{\label{tab:table1} Experimental conditions for Fig.~\ref{fig7_spin_precession}.}
\begin{tabular}{c|c}
\hline
$N$&2447514\\
\hline
$f_\mathrm{rep1}$  &61,530,120.000 Hz\\ \hline
$f_\mathrm{rep2}$  &61,530,145.140 Hz\\ \hline
$\Delta f_\mathrm{rep}$  &25.140 Hz\\ \hline
 $\tau_\mathrm{step}$  &6.640 fs\\ \hline
 $\tau_\mathrm{total}$  &16.252 ns\\ \hline
 $\tau_\mathrm{scan}$  &39.777 ms\\ \hline
  $H_\mathrm{ext}$  & 417 mT\\ \hline
\end{tabular}
\end{table}

\subsection{Magnetic field dependence of the spin precession frequency}

\begin{table}
\caption{\label{tab:table2} Experimental conditions for Fig.~\ref{fig8_Hext_dependance}.}
\begin{tabular}{c|c}
\hline
$N$&162976\\
\hline
$f_\mathrm{rep1}$  &61,530,120.000 Hz\\ \hline
$f_\mathrm{rep2}$  &61,530,497.541 Hz\\ \hline
$\Delta f_\mathrm{rep}$  &377.541 Hz\\ \hline
 $\tau_\mathrm{step}$  &99.721 fs\\ \hline
 $\tau_\mathrm{total}$  &16.252 ns\\ \hline
 $\tau_\mathrm{scan}$  &2.649 ms\\ \hline
\end{tabular}
\end{table}

\begin{figure}
\includegraphics[width=0.5\linewidth]{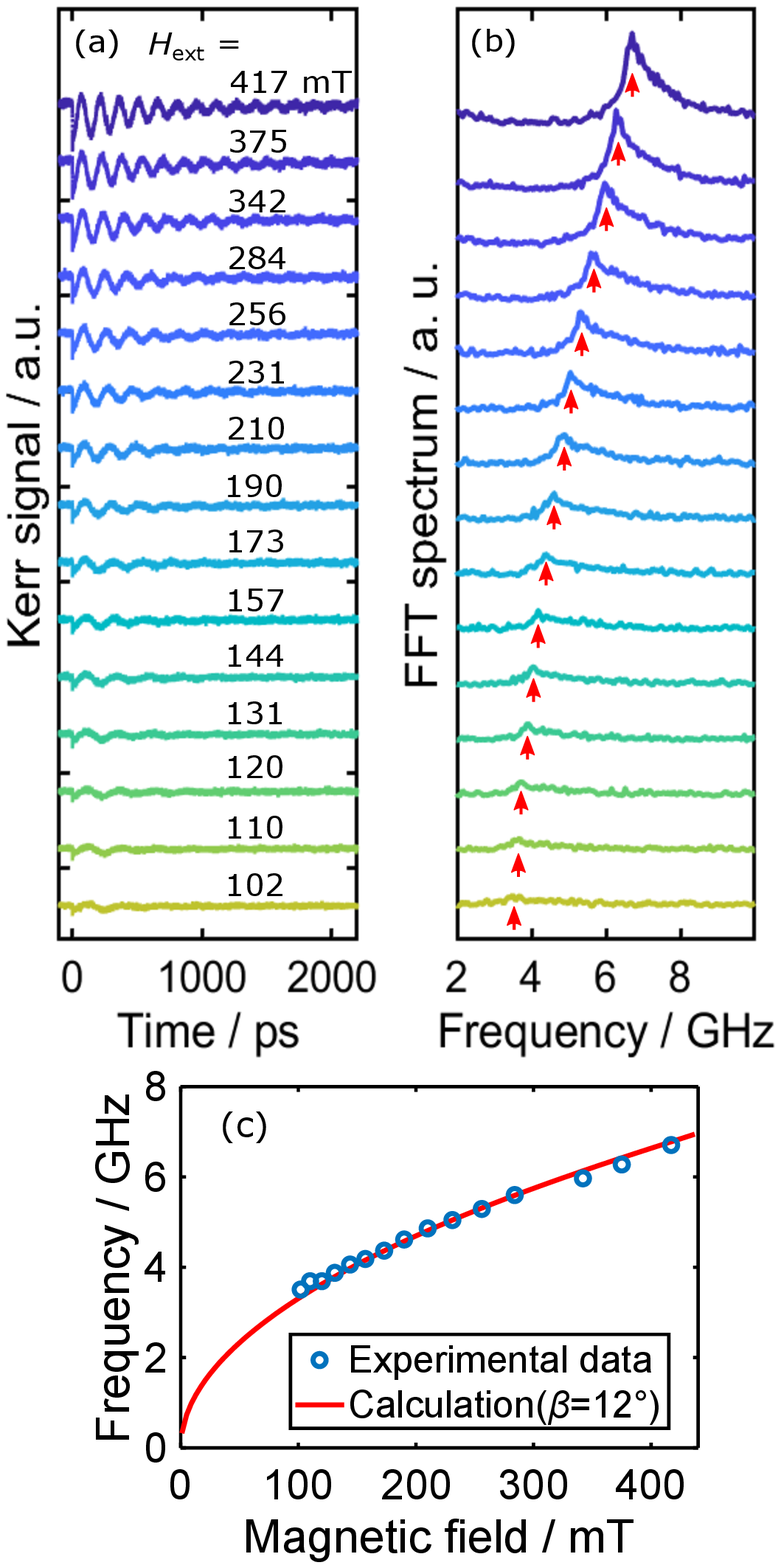}
\caption{\label{fig8_Hext_dependance} (a) The dependence of the resonance frequency on the external magnetic field. (b) The corresponding frequency-domain waveforms obtained by FFT. The red arrows indicate the peak positions. (c) Comparison between the numerical prediction based on the Smit–Beljers approach and the experimental values ($\beta = 12^\circ$).}
\end{figure}

In this section, we evaluate the magnetic field dependence of the spin precession frequency and verify the validity of our measurement procedure. Table~\ref{tab:table2} shows the measurement conditions for these experiments. We set $\Delta f_\mathrm{rep}\approx377.541$ Hz for a faster evaluation of the spin precession frequency, and therefore the time resolution is determined by the detector bandwidth; $t_\mathrm{det} = $6.1 ps, which is sufficiently short to trace the oscillation. Figure~\ref{fig8_Hext_dependance}(a) shows the TR-MOKE signals obtained from the sample in different external magnetic fields. For each curve, we accumulated the TR-MOKE waveform about 5000 times, which takes about 13 seconds. Figure~\ref{fig8_Hext_dependance}(b) shows the spectra obtained by FFT of the curves in Fig.~\ref{fig8_Hext_dependance}(a), and we can confirm that the peak frequency of the spin precession motion increases with the external magnetic field. Figure~\ref{fig8_Hext_dependance}(c) plots the peak frequency as a function of the external magnetic field. This experimental result can be described by using the approach of Smit and Beljers as follows~\cite{Smit1955}: First, in the numerical treatment of the Landau–Lifshitz–Gilbert (LLG) equation, the zenith angle of the magnetization of the Py thin film, denoted by $\theta$, can be derived from the following equation as a function of $H_\mathrm{ext}$:
\begin{equation}
    2 H_\mathrm{ext} \sin{(\theta-\beta)} - M_s \sin{2\theta} = 0,
\end{equation}
where $M_s = 522000$ $\mathrm{A \cdot m^{-1}}$ is the saturation magnetization of Py and $\beta$ is the zenith angle of the external magnetic field, which is 12$^\circ$ in our measurement. Then, the resonance frequency of the spin precession is given by~\cite{Smit1955,PhysRevB.97.014438},
\begin{equation}
    f = \frac{\gamma \mu_0}{2 \pi} \sqrt{ [{H_\mathrm{ext} \cos{(\theta-\beta) - M_s \cos{2 \theta}}]} \times {[H_\mathrm{ext} \cos{(\theta-\beta) - M_s \cos^2{ \theta}}}] }, \label{eq:f}
\end{equation}
where $\gamma=1.76252\times 10^{11}~\mathrm{s^{-1}\cdot T^{-1}}$ is the gyromagnetic ratio.  The solid curve in Fig.~\ref{fig8_Hext_dependance}(c) represents the theoretical value of $f$ as a function of $H_\mathrm{ext}$ derived from Eq.~(\ref{eq:f}). The experimental results are in good agreement with the theoretical prediction, which indicates that our measurement method was successfully applied to the characterization of optically induced spin precession.

\section{Conclusion}
We have demonstrated time-resolved measurements of ultrafast magnetization dynamics in a ferromagnetic thin film using a triggerless ASOPS system. The light source for the pump–probe measurements was a two-color dual-comb system, and we acquired time-resolved signals in the range from picoseconds to nanoseconds within relatively short measurement times (a few tens of seconds including the accumulation of the data). We have shown that the observed external magnetic field dependence of the spin precession frequency of the Py thin film agrees well with the numerical prediction. We consider that this ASOPS system based on the dual-comb architecture enables us to perform high-resolution and ultrafast measurements of magnetization dynamics. Furthermore, this system can be a powerful tool for experiments that require many individual measurements, such as imaging experiments (for example, we consider the application of such an ASOPS system to spatio-temporal imaging of spin wave propagation, which is an extension of the present study). The measurement time of this method is much shorter than that of the conventional pump–probe method, which is based on using a delay stage, and there are many potential applications of such a method for the rapid characterization of magnetization dynamics in optically pumped magnetic samples. 

\section*{Acknowledgments}
We would like to acknowledge Prof. Yukio Nozaki for fruitful discussions and thank Ms. Imai for her help in constructing the experimental system. This work was partially supported by JST, CREST Grant number JPMJCR19J4, and MEXT Quantum Leap Flagship Program (MEXT Q-LEAP), Grant No. JPMXS0118067246.


%
%

%


\bibliography{Nishikawa2022_rsi}
\bibliographystyle{apsrev4-1}

\end{document}